\documentclass[11pt,twoside]{article}
\usepackage{subfig}

\usepackage{graphicx}
\usepackage{url}

\begin{document}

\begin{center}
{\Large \bf Polarisation and Polarimetry at HERA}\\
\vspace{0.8cm}
{\large Blanka Sobloher\\for the POL2000 collaboration}
\vspace{0.5cm}
{\it Deutsches Elektronen--Synchrotron DESY\\
Notkestr. 85, 22607 Hamburg, Germany\\
E-mail: blanka.sobloher@desy.de}\\
\vspace{0.8cm}
{\it Proceedings prepared for the}\\
 13th International Workshop On Polarized Sources And Targets \& Polarimetry (PST 2009), 7-11 Sep 2009, Ferrara, Italy\\
\vspace{0.5cm}
\end{center}

\begin{abstract}
Longitudinal polarisation of the lepton beam is a key ingredient to the success of the world's unique $e^\pm p$ ring collider HERA. This article aims at
providing a brief introduction to the physics motivation for deep-inelastic scattering of polarised electrons or positrons off protons, the basic
mechanisms to establish lepton polarisation in the high-energy storage ring and to describe briefly the three different polarimeters, which measured both
the transverse and the longitudinal polarisation.
\end{abstract}


\section{Introduction}
The unique HERA facility in Hamburg, Germany collided leptons with protons at centre-of-mass energies of $300$ and $318\,\mathrm{GeV}$ between 1991
and 2007, incorporating radiative polarisation of the lepton beam.
Spin rotators installed around the interaction points of the experiments HERMES, H1 and ZEUS transformed the natural transverse polarisation of the
lepton beam to logitudinal polarisation, which is in deep-inelastic scattering a powerful tool to study the internal structure of the nucleus.

The polarisation was measured routinely with two polarimeters.
Using polarisation dependent Compton scattering, the transverse polarimeter TPOL detected the tiny up--down asymmetries associated with vertical
polarisation, while the longitudinal polarimeter LPOL utilised the energy asymmetry caused by longitudinal polarisation.
The third polarimeter, employing a Fabry--Perot cavity to provide a high laser photon density to measure longitudinal polarisation,
started to collect significant amounts of data towards the end of 2006 and in 2007.

\section{The HERA Collider}
\begin{figure}[ht]
  \centering
     \includegraphics[width=2.5in]{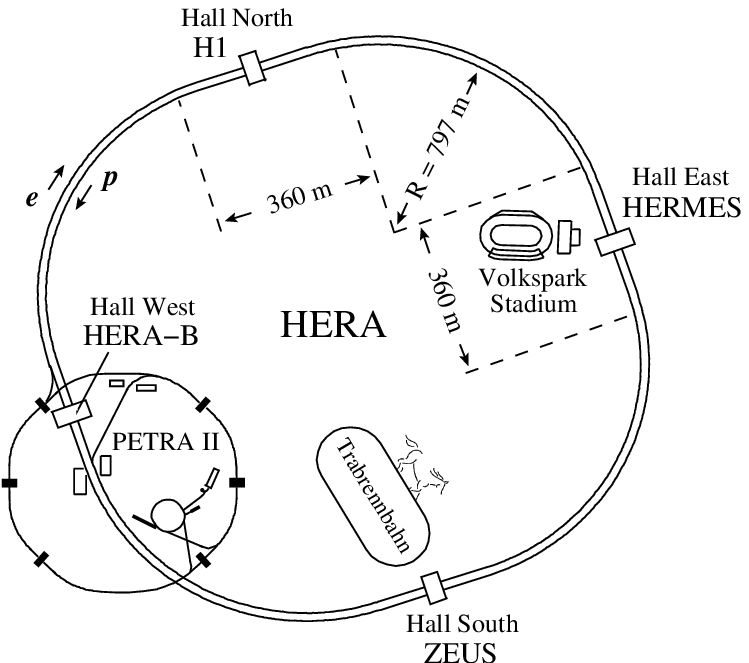}
    \hfill
     \includegraphics[width=2.5in]{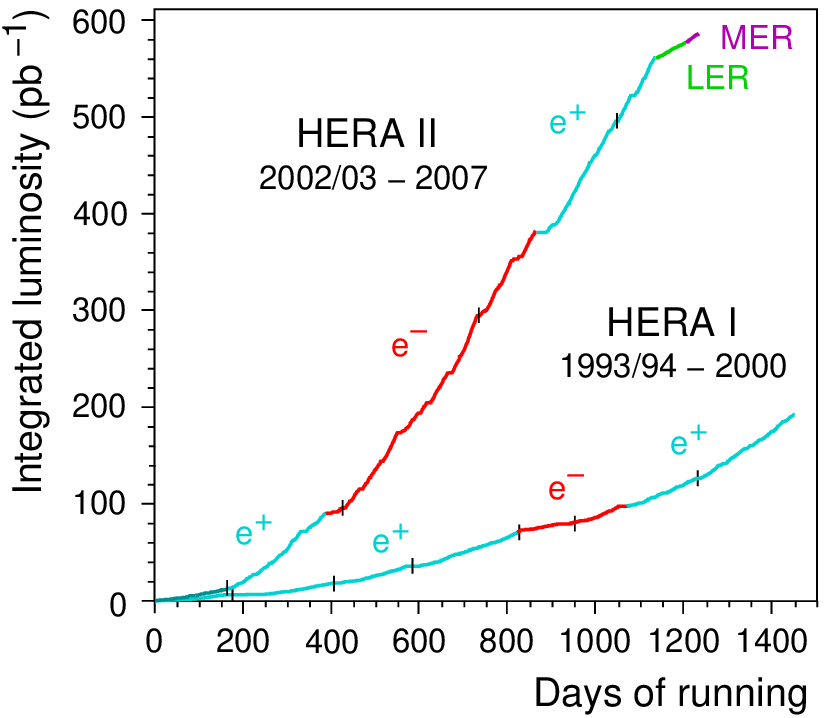}
   \caption{The HERA accelerator complex (left) and the integrated luminosity at HERA I and HERA II
     (right)\cite{lumi}. Labels LER and MER: Low and Middle (proton) Energy Runs, black marks:
     approximate change of year.} 
   \label{fig:hera}
\end{figure}

The \emph{Hadron Elektron Ring Anlage} HERA is the first and only electron--proton or positron--proton storage ring, located at the
\emph{Deutsches Elektronen--Synchrotron} DESY laboratory in Hamburg, Germany. First $e^\pm p$ collisions were achieved in October 1991, with the
colliding beams experiments ZEUS and H1 taking first physics data shortly afterwards.\cite{Voss:1994} The fixed target experiments HERMES and HERA--B
went into operation in 1995 and 2000 respectively, the latter taking data till March 2003. The collider went through an ambitious luminosity upgrade
in 2000/01 and has been shut down finally after a successfull second running period on 1 July 2007. A schematic of the HERA accelerator complex is
shown in Fig.~\ref{fig:hera}~(left).

At HERA electron or positron beams were accelerated to the energy of $27.5\,\mathrm{GeV}$ and the counter-rotating proton beams to energies of
$820\,\mathrm{GeV}$ or $920\,\mathrm{GeV}$. Colliding at the interaction points of ZEUS and H1 the beams yielded a centre-of-mass
energy an order of magnitude larger than conventional fixed target experiments. 
From the beginning HERA was designed to incorporate a polarised lepton beam and the establishment of transverse polarisation in the storage ring
was an important prerequisite for the HERMES experiment. Longitudinal polarisation was delivered to HERMES from its
beginning and since the upgrade also to H1 and ZEUS. Till the final shutdown each of the colliding beams experiments collected an
integrated luminosity of $\approx 0.5\,\mathrm{fb}^{-1}$ as can be taken from Fig.~\ref{fig:hera}~(right).

\subsection{Physics Case for longitudinal Polarisation}
At HERMES, the longitudinal polarised lepton beams scattered off gas targets to study the nucleon spin structure by determining the spin-dependent
structure functions of various nuclei. The helicity distributions of individual quark flavours inside the nucleon could be determined and further
information on the spin structure of the nucleon has been obtained by investigating whether the gluons inside of the nucleon are also
polarised.\cite{Steenhoven:2005}

Being delivered with a longitudinally polarised lepton beam, the physics programs of H1 and ZEUS were extended by a large electroweak program,
allowing to study for instance the chirality of charged current (CC) interactions. The CC deep-inelasting scattering (DIS) cross section
depends on the polarisation of the lepton colliding with the proton.\cite{H1ew:2006,Glazov:2009}
According to the Standard Model only left handed fermions couple to the $W$--boson, the cross section should therefor vanish for fully right-handed
polarised electrons and fully left-handed positrons. By setting an upper limit on a non-vanishing cross section a lower limit on the mass of a
hypothetical right-handed $W$--boson can be set.

There is also sensitivity to some electroweak parameters of the Standard Model like the $W$--boson mass $M_W$. The ratio of neutral
current (NC) to CC cross sections constrains the $W$--boson mass in the $(M_W,M_t)$ plane. 
The sensitivity to the $W$--boson mass and the electroweak mixing angle $\sin^2 \theta_W = 1 - M_W^2/M_Z^2$ provides also a test for electroweak
universality.\cite{Beyer:1995}

Other examples are given by the measurement of the light quark ($u$, $d$) neutral current couplings and the $\gamma Z^0$ interference structure
functions $F_2$ and $xF_3$ by detailed comparison of polarised NC and CC cross sections\cite{Klein:2003} and 
potential for new physics in e.g.~searches for leptoquarks or R--parity violating supersymmetry.

\subsection{Radiative Polarisation}
Acceleration in a circular machine like HERA implies crossing of many depolarising resonances, making it difficult to sustain an initial polarisation
at a high degree.\cite{Steffens:2008}
Instead, the inevitable emission of synchrotron radiation is used to polarise the stored beam after acceleration has finished. By means of spin-flips
caused by a small fraction of synchrotron emissions in the magnetic field of bending dipoles, the spin of emitting particles align parallel or
anti-parallel with the transverse magnetic field, leading to a gradual build-up of polarisation.
This \emph{radiative polarisation} has been first described by Sokolov and Ternov in 1964.\cite{Sokolov:1963zn}

In an ideal machine the build-up of radiative polarisation $P$ proceeds exponentially with 
\begin{equation}
   P(t) = P_\mathrm{ST} \bigl( 1-e^{-t/\tau_\mathrm{ST}} \bigr)
\end{equation}
with the asymptotic polarisation limit and build-up time given by \mbox{$P_\mathrm{ST} = 8/(5\sqrt{3})\approx0.924$} and
$\tau_\mathrm{ST} \approx 100\,\mathrm{s} \cdot \rho^3/E^5 \cdot \mathrm{GeV}^5/\mathrm{m}^3$.

In a real storage ring the polarisation build-up is counteracted by several depolarising effects, causing the maximal achievable polarisation $P$ and
build-up time $\tau$ to be smaller than given by the Sokolov--Ternov effect alone.
Spin diffusion in the presence of misalignments, field errors and horizontal magnetic fields along the ring weaken the polarisation build-up and   
careful alignment and organisation of the quadrupole strengths, known as \emph{spin matching}, along with \emph{harmonic orbit corrections} are
needed to minimise the influence.\cite{Barber:1994}
In addition, depolarising resonances are avioded by choosing a half-integer spin-tune, i.e.~the number of precessions a spin performs per turn in the
ring $\nu \mathrel{\mathop:}= a \gamma$ with $a=(g-2)/2$ being the electron gyromagnetic anomaly. HERA operated at $E=27.5\,\mathrm{GeV}$ had a
spin-tune of $\nu=62.5$. 

\subsection{Spin Rotators}
In order to obtain longitudinal polarisation at the interaction points of the experiments, pairs of spin rotators were installed around each interaction
region, exploiting spin precession arising from deflection in transverse magnetic fields. As the spin-tune $\nu$ is large, small commuting orbit
deflections $\phi$ can be used to generate large non-commuting spin precessions $\psi=\nu\phi$.

At HERA the socalled \emph{mini--rotator} design of Steffen and Buon has been adopted, consisting of a series of six alternating vertical and
horizontal bends without any focussing quadrupoles within due to its relatively small length of $56\,\mathrm{m}$.\cite{Buon:1985de} These dipole spin
rotators allow to have either sign of electron helicity in the longitudinal spin state, though with a maximum orbit distortion of $\pm22\,\mathrm{cm}$
the bends had to be moved vertically upon a helicity change.
The first rotator pair has been installed in 1993/94 for the HERMES experiment, followed by further pairs for H1 and ZEUS during the HERA
upgrade in 2000/01.

\subsection{Polarisation at HERA}
A typical fill of the HERA electron ring could last for more than 12 hours. Initially filled with currents of about $40\,\mathrm{mA}$ in $180-190$
bunches with a spacing of $96\,\mathrm{ns}$, the bunch current decreased during a fill over time due to collisions and other losses. 

The polarisation at HERA was monitored independently by two fast polarimeters with a very high availability and providing realtime polarisation values
to the machine and the experiments. During the HERA~II running period over $99\,\%$ of all physics fills had at least one polarimeter operational. 
The maximum polarisation ever achieved was about $0.76$ in the HERA~I period before the upgrade. 
As spin rotators represent a source of spin diffusion, the typical equilibrium polarisation at HERA~II with three spin rotator pairs was lower
with $\approx0.4-0.5$ and rise times about $40\,\mathrm{min}$. The polarisation varied from fill to fill and even within a single fill, as it was
subject to the tuning of the machine. In addition, colliding and non-colliding bunches had different asymptotic polarisation values due to beam tune
shifts of the colliding bunches caused by beam--beam interactions with the proton bunches.

\section{The HERA Polarimeters}
The requirements of polarimetry at HERA were challenging. 
The polarimeters should measure the polarisation of the stored lepton beam continuously in a non-invasive manner, providing realtime values
to the experiments and the HERA machine control with a statistical accuracy of few percent per minute measurement.
In addition, the devices required the ability to handle the frequent changes in the lepton orbit. The systematical uncertainties should be small with
\mbox{$\delta P / P < 2.5\,\%$}, if measurements relying on polarisation values shall not be dominated by polarimetry.

One polarimeter measured the transverse polarisation and two more measured the longitudinal polarisation within the HERMES straight section.
The basis for all three devices is given by Compton scattering laser photons of the high-energetic lepton beam and
observing the backscattered photons. The cross section for Compton scattering is sensitive to the transverse and longitudinal components $P_y$ and
$P_z$ of the lepton beam polarisation, provided that the laser photons are circularly polarised:
\begin{equation}
  \frac{d^2\sigma}{dEd\phi} = \Sigma_0(E) + S_1 \Sigma_1(E)\cos 2\phi + S_3 P_y \Sigma_{2y}(E) \sin \phi + S_3 P_z \Sigma_{2z} (E)
\end{equation}
with $S_1$ and $S_3$ being the linear and circular components of the laser light polarisation.\cite{Lipps:1954a}
The polarisation is then measured using the asymmetry of the cross sections, when switching the laser polarisation between left and right
helicity states, i.e.~$S_3=\pm1$ with $S_1\approx0$:
\begin{equation}
   \mathcal{A}(y,E_\gamma) = \frac{\sigma_L(y,E_\gamma) - \sigma_R(y,E_\gamma)}{\sigma_L(y,E_\gamma) - \sigma_R(y,E_\gamma)}
\end{equation}

\subsection{Transverse Polarimeter TPOL}
The transverse polarimeter (TPOL) operated throughout the complete HERA~I and II periods in the straight section West near the HERA--B experiment.
It measured the tiny spatial asymmetry between the left and right laser helicity states caused by the transverse polarisation.\cite{Barber:1993,Barber:1994}

A green Argon--Ion laser, operated at $10\,\mathrm{W}$ in cw mode, was made circularly polarised by means of a Pockels cell, switching the helicity at
a frequency of $\approx80\,\mathrm{Hz}$. The laser beam was transported by an optical system over more than $300\,\mathrm{m}$ into the HERA tunnel and was
brought into collision with the lepton beam under a vertical angle of $3.1\,\mathrm{mrad}$. The degree of light polarisation was regularly monitored
behind the Compton interaction point using a rotating Glan prism, with typical polarisation values $S_3>0.99$.

The backscattered Compton photons were detected $65\,\mathrm{m}$ downstream of the Compton interaction point in a compact, $19\,X_0$ deep
electromagnetic scintillator--tungsten sampling calorimeter, read out using wavelength shifter bars from all four transverse sides. 
To achieve sensitivity to the vertical position of the incident photon, the scintillator plates were optically decoupled along the central horizontal
plane, thus dividing the calorimeter effectively into independent upper and lower halves. 
Information about the energy and the vertical impact position $y$ of an incident photon is then obtained from the sum of the two halves \mbox{$E =
  E_\mathrm{up} + E_\mathrm{down}$} and the energy asymmetry $\eta$ between them:
\begin{equation}
    \eta \mathrel{\mathop:}= \frac{E_\mathrm{up} - E_\mathrm{down}}{E_\mathrm{up} + E_\mathrm{down}}
\end{equation}

The operation of the transverse polarimeter relied on the \emph{single--photon mode} with on average $\bar{n}=0.01$ backscattered photons per bunch
crossing, so that, if at all, only one photon will be detected.
While this requires relatively low photon rates of $<100\,\mathrm{kHz}$, it allows to use the known kinematical endpoint of the Compton scattering process,
called \emph{Compton edge}, for the absolute calibration of the detector.
The main background is bremsstrahlung generated along the $7.3\,\mathrm{m}$ short straight section which is in the line of sight of the detector.
The measurement of the backscattered photon distributions separately for the two laser helicity states was interspersed regularly with
measurements where the laser was blocked off. This allowed to measure the background and to subtract it from the Compton data on a statistical basis.

The polarisation of colliding and non-colliding bunches is measured separately and since the upgrade in 2000/01 also a bunchwise measurement is made
possible by means of a new faster DAQ. During this upgrade also a position sensitive detector in the form of crossed silicon strip detectors including the
necessary preradiator of $1\,X_0$ thickness have been added in front of the calorimeter. These detectors should allow for an \emph{in situ} measurement of the
intrinsic calorimeter response, the non-linear transformation between the spatial impact point $y$ and the energy asymmetry $\eta$ called
\emph{$\eta(y)$--transformation}.

\begin{figure}[htb]
   \centering
   \includegraphics[width=2.5in]{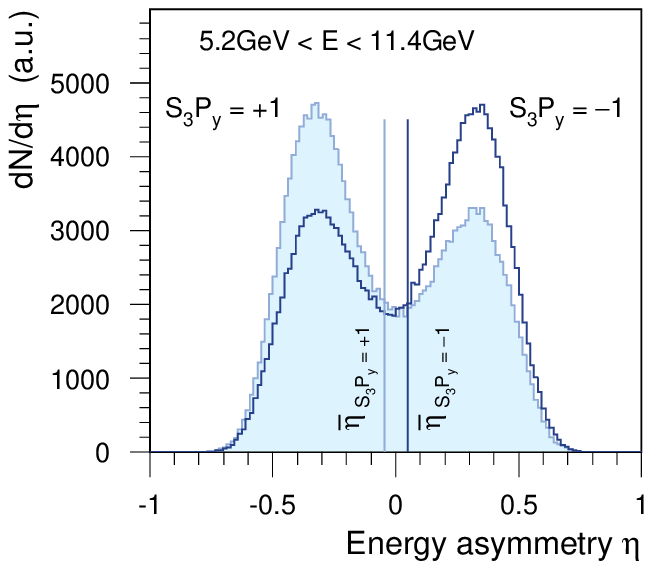}
   \hfill
   \small{
   \begin{tabular}[b]{lc}\hline
       Systematic uncertainty & $\Delta P/P$ \\\hline \hline
       Electronic noise & $<\pm0.1\,\%$ \\      
       Calorimeter calibration & $<\pm0.1\,\%$ \\
       Background subtraction & $<\pm0.1\,\%$ \\
       Laser light polarisation & $\pm0.1\,\%$ \\
       Compton beam centering & $\pm0.4\,\%$ \\
       Focus correction & $\pm1.0\,\%$ \\
       Interaction point region & $\pm0.3\,\%$ \\
       Interaction point distance & $\pm2.1\,\%$ \\
       Absolute scale & $\pm1.7\,\%$ \\ \hline
       Total syst. uncertainty & $\pm2.9\,\%$ \\ \hline \\
  \end{tabular}}
  \caption{Illustration of the polarisation dependent shift of the mean energy asymmetry
    distributions (left) and the preliminary list of contributions to the fractional systematic
     uncertainty of the TPOL measurement (right).}\label{fig:tpol_polandsys}
\end{figure}

The polarisation is then calculated from the shift of the mean energy asymmetry distributions for left and right laser helicity states using an
analysing power $\Pi$: 
\begin{equation}
  \bar{\eta}_\mathrm{L} - \bar{\eta}_\mathrm{R} \mathrel{\mathop:}= \Delta S_3 P_y \Pi
\end{equation}
as illustrated in Fig.~\ref{fig:tpol_polandsys}~(left).
At HERA~I the analysing power was determined from simulations and from rise time measurements in a flat machine, where the intrinsic relation
between the asymptotic polarisation value and the rise time constant as given by the Sokolov--Ternov effect is exploited for the calibration of the
absolute polarisation scale. 
At HERA~II the beam conditions as well as the detector have changed. Both the lepton beam size and divergence as well as the
longitudinal position of the Compton interaction point became more variable, influencing the photon distribution at the calorimeter surface and thus
the analysing power. Also the exchange of the calorimeter and the added dead material in front are likely to change the analysing power with respect
to the HERA~I running period.

The statistical uncertainty of the polarisation measurement amounts to about $2-3\,\%$ per minute of data, for single bunches to about $10\,\%$ per
10 minutes of measurement. The current, preliminary estimation of systematic uncertainties amounts to $2.9\,\%$ as is shown in the table in
Fig.~\ref{fig:tpol_polandsys}~(right).\cite{Airapetian:2007}
The dominant contribution is given by the analysing power $\Pi$, in the current breakdown of sources divided into three contributions given by the
influence of the intrinsic beam width and divergence, the distance of the Compton interaction point and the absolute scale. While
the first of the three has been corrected for since 2004\cite{Corriveau:2004}, for the second only an upper limit from geometrical
acceptances is known. The three mentioned dominant contributions are correlated and have to be evaluated thus in a correlated fashion using a detailed
realistic simulation of the magnetic beam line and a precise modelling of the calorimeter response including $\eta(y)$--transformation and energy resolution.

\subsection{Longitudinal Polarimeter LPOL}
The second polarimeter (LPOL) measured the longitudinal lepton beam polarisation within the HERMES spin rotator pair, downstream of the HERMES gas
target.\cite{Beckmann:2000} It went into operation in 1997 and used the sizable asymmetries in the energy distributions of the backscattered Compton
photons when switching between the left and right laser helicity states.

The polarimeter operated in \emph{multi--photon mode}, where on average $\bar{n}\approx10^3$ photons are backscattered per bunch crossing. In
this mode background like bremsstrahlung becomes less important. With most of the long straight section East in the line of sight of the calorimeter,
bremsstrahlung background would be too high to operate in single--photon mode.

The key ingredient to such high backscattering probabilities are high power laser pulses. Generated by a frequency-doubled
green Nd:YAG laser, pulsed at $100\,\mathrm{Hz}$, each laser pulse had a fixed power of  $100\,\mathrm{mJ}$ and a length of $3\,\mathrm{ns}$.
The laser was synchronized with the lepton bunches and a trigger for readout at twice the laser pulse frequency allowed
to measure background every second event. The circular polarisation, achieved by a Pockels cell flipping helicity for every pulse, was measured using a
Glan--Thompson prism regularly with $S_3>0.99$.

The laser was transported with an optical system over $70\,\mathrm{m}$ into the HERA tunnel and collided with the lepton
beam at a vertical crossing angle of $8.7\,\mathrm{mrad}$. The backscattered Compton photons were detected $54\,\mathrm{m}$ downstream by a compact
electromagnetic \v{C}erenkov calorimeter, consisting of four $19\,X_0$ deep NaBi(WO$_4$)$_2$ crystals (NBW), read out separately.
The crystals were optically decoupled and arranged in a rectangular $2\times2$ array to allow for a positioning of the calorimeter in
the photon beam.

\begin{figure}[htb]
  \centering
   \includegraphics[width=2.5in]{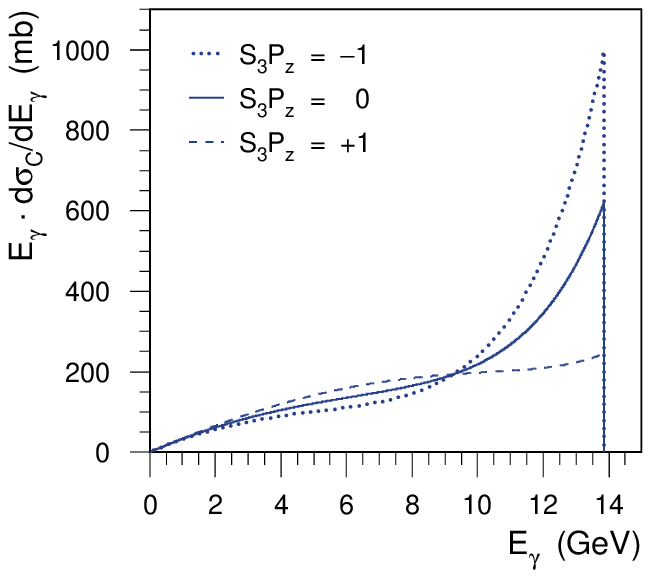}
   \hfill
   \small{
\begin{tabular}[b]{llcc}\hline
       \multicolumn{2}{l}{Systematic uncertainty}       & \multicolumn{2}{c}{$\Delta P/P$} \\\hline
       \multicolumn{2}{l}{Analysing power}                                 & \multicolumn{2}{c}{$\pm1.2\,\%$} \\
       \multicolumn{2}{l}{\quad long-term stability}    & \multicolumn{2}{c}{$\pm0.5\,\%$} \\
       \multicolumn{2}{l}{Gain matching}                                     & \multicolumn{2}{c}{$\pm0.3\,\%$} \\
       \multicolumn{2}{l}{Laser light polarisation}                        & \multicolumn{2}{c}{$\pm0.2\,\%$} \\
       \multicolumn{2}{l}{Helicity dep. luminosity}              & \multicolumn{2}{c}{$\pm0.4\,\%$} \\
       \multicolumn{2}{l}{Interaction region stability}                   & \multicolumn{2}{c}{$\pm0.8\,\%$} \\
       \multicolumn{2}{l}{Total (HERA I)}   & \multicolumn{2}{c}{$\pm1.6\,\%$} \\\hline
       \multicolumn{2}{l}{Extra (new calorimeter)}   & \multicolumn{2}{c}{$\leq \pm1.2\,\%$} \\\hline
       \multicolumn{2}{l}{Total (HERA II)}  & \multicolumn{2}{c}{$\pm2.0\,\%$} \\\hline\\
\end{tabular}}
   \caption{The energy-weighted single differential Compton cross section $E_\gamma d\sigma_\mathrm{C} / dE_\gamma$ (left) and the list of
     contributions to the fractional systematic uncertainty of the LPOL measurement (right).}\label{fig:lpol_polandsys}
\end{figure}

In the multi--photon mode, the detector signal is proportional to the integral of the energy-weighted Compton cross section:
\begin{equation}
   I_{S_3 P_z} \mathrel{\mathop:}= \int_{E_\gamma^\mathrm{min}}^{E_\gamma^\mathrm{max}} r(E_\gamma) E_\gamma \frac{d\sigma_\mathrm{C}}{dE_\gamma}
   dE_\gamma 
\end{equation}
with $r(E_\gamma)$ being the single--photon relative response function, a constant for a perfectly linear detector. The energy-weighted Compton cross
section is shown in Fig.~\ref{fig:lpol_polandsys}~(left). The energy dependent asymmetry then becomes
\begin{equation}
   \mathcal{A} \mathrel{\mathop:}= \frac{I_{S_3 P_z <0} - I_{S_3 P_z >0}}{I_{S_3 P_z <0} + I_{S_3 P_z >0}} = \Delta S_3 P_z \Pi_z
\end{equation}
The statistical uncertainty of the measurement is about $1-2\,\%$ per minute and about $6\,\%$ per 5 minutes measurement, clearly limited by the
repetition rate of the laser.

The current estimation of systematic uncertainties is shown in the table in Fig.~\ref{fig:lpol_polandsys}~(right).\cite{Airapetian:2007} The dominant
systematic uncertainty is given by the analysing power $\Pi_z = 0.1929\pm0.0017$.\cite{Airapetian:2005} Its main contributions are given by the shape
of the single--photon response function as measured with test beam data and the extrapolation from single-- to multi--photon mode. The latter was validated
by attenuating the signal over three orders of magnitude using neutral density filters and monitoring the polarisation value in comparison with the
independent measurement of the TPOL.
After the replacement of the calorimeter crystals in 2004 the performance of the new calorimeter was ascertained in alternating measurements with a
sampling calorimeter. From this an upper limit of $1.2\,\%$ systematic uncertainty due to the new calorimeter was estimated, increasing the formerly
quoted HERA~I systematic uncertainty to $2\,\%$.\cite{Airapetian:2007}

\subsection{Cavity Longitudinal Polarimeter}
A third polarimeter project has been started in the early HERA~II running phase employing a Fabry--Perot cavity to stock laser photons with a very high
density at the Compton interaction point. Working in continuous \emph{few--photon mode}, backscattering on average $\bar{n}\approx1$ photons per
bunch crossing, it combines the virtues of both existing operational methods. While providing a very high statistics with
scattering rates in the order of MHz, it can make use of the Compton and bremsstrahlung edges for the calibration of the calorimeter.

The cavity polarimeter measured the longitudinal polarisation within the HERMES spin rotator pair, located about $10\,\mathrm{m}$ downstream of the
LPOL interaction point and utilising the same detector location for the measurement of the backscattered Compton photons. After installation of the
Fabry--Perot cavity in spring 2003, first Compton events have been observed in March 2005 with a much increased operation till the end of HERA. Over
500 hours of efficient data could be collected.

The cavity is driven by an infrared Nd:YAG laser with an intial power of $0.7\,\mathrm{W}$, located together with all optical components on an optical
table close to the cavity. Circular polarisation of the laser light is achieved by rotating quarter wave plates, flipping the helicity every few seconds, and
monitored behind the cavity.\cite{Zhang:2001}
The cavity mirrors are located inside the vacuum vessel at $2\,\mathrm{m}$ distance from each other, providing a vertical crossing angle of $3.3^\circ$.
With a finesse of $\approx3\times10^4$ the initial laser power is amplified by means of constructive interference with an effective gain of
$\approx5000$ to about $3\,\mathrm{kW}$.\cite{Zomer:2003}

The measurement of the longitudinal polarisation proceeds by an overall fit of a parametrised model to the energy distributions for the two laser
helicity states collected separately. Absolute calibration is done using the known Compton and bremsstrahlung edge positions. The description of the
energy spectra includes besides the Compton spectrum also contributions of background like synchrotron and Compton scattered black-body radiation, the
bremsstrahlung spectrum as well as detector resolution and non-linearity parameters. Detailed simulations of the calorimeter response
were needed, e.g~for a precise description of the synchrotron radiation peak.
The statistical uncertainty with about $3\,\%$ per bunch and $10\,\mathrm{s}$ doublet is unprecedented at HERA.

Based on more than 500 hours of data including dedicated data samples, most of it taken during the final stage of the HERA operation, detailed
systematic studies have been performed.
The preliminary list of systematic uncertainties includes the modelling of the detector response and of the synchrotron radiation peak, electronic
pile-up, detector parameter fitting, a varying HERA beam, the calorimeter position and the laser polarisation inside the cavity. Whereas the
contributions of parameter fitting and HERA beam variations are found to be negligible, the other contributions are of approximately the same size,
adding to a total of $\delta P/P=0.9\,\%$.\cite{Cavity:09}

\section{Conclusions}
The running of HERA was efficiently covered with measurements of the lepton beam polarisation. At HERA~II over $99\,\%$ of all the physics fills had
at least one polarimeter operational. 

The preliminary estimation of the systematical uncertainties for TPOL amounts to about $2.9\,\%$ and for LPOL to $2\,\%$. However, the agreement of
the two polarimeters shows a varying behaviour over time which is not yet understood. To cover these discrepancies an additional systematical
uncertainty of $3\,\%$ had been assigned, raising the uncertainty of the combined measurement to about $3.4\,\%$.\cite{Airapetian:2007} Currently,
efforts are under way to validate and improve the polarisation analyses of both polarimeters to decrease the systematical uncertainty of the combined
measurement and final results are expected within the next few months.

The polarisation measurement with a high finesse Fabry--Perot cavity at HERA has been established, successfully operating with increasing data taking
frequency till the end of HERA. The analysis of the systematical studies is nearly finished, indicating that the goal of a sub percent systematic
precision has been achieved.

\section*{Acknowledgments}
Special thanks are due to T.~Behnke, R.~Fabbri and Z.~Zhang giving me invaluable advise for the preparation of the talk and this article. I am
particularly indebted to the organisers of the PST2009 conference, whose support made my contribution possible.

\begin{footnotesize}
\bibliographystyle{num_thesis}
\bibliography{proceedings_sobloher}
$^*$see also \url{http://www.desy.de/~pol2000}
\end{footnotesize}
\end{document}